\documentclass[runningheads,envcountsame]{llncs}

\makeatletter \RequirePackage[bookmarks,unicode,colorlinks=true]{hyperref}%
\def\@citecolor{blue}%
\def\@urlcolor{blue}%
\def\@linkcolor{blue}%

\def\orcidID#1{\href{http://orcid.org/#1}{\smash{\protect\raisebox{-1.25pt}{\protect\includegraphics{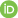}}}}}
\makeatother

\usepackage[hyperref=true, backend=bibtex, firstinits=true, maxbibnames=99, sortcites, style=numeric-comp]{biblatex}
\usepackage{mathtools,amsfonts,amsmath,amssymb}
\usepackage{tikz}
\usetikzlibrary{shapes,calc,arrows,automata,decorations.pathmorphing,decorations.pathreplacing,positioning,positioning,fit,calc,backgrounds}

\usepackage{hyperref}
\hypersetup{
  pdftitle={KoAT: Automatic Complexity and Termination Analysis of Integer Programs}, colorlinks=true, linkcolor=blue, citecolor=olive, filecolor=magenta, urlcolor=cyan
}

\usepackage[dvipsnames]{xcolor}
\definecolor{burgundy}{HTML}{AC4142}

\usepackage[capitalize,nameinlink]{cleveref}
\usepackage{todonotes}
\setuptodonotes{inline}

\usepackage{multicol}
\usepackage{hyperref}
\usepackage{tabto}
\usepackage{thm-restate}
\usepackage{stmaryrd}
\usepackage{booktabs}

\usepackage[shortlabels,inline]{enumitem}
\setlist[enumerate,1]{label=(\alph*), wide=0pt, leftmargin=*}
\setlist[itemize,1]{label=\textbullet}

\makeatletter
\newcommand{\inlineitem}[1][]{%
  \ifnum\enit@type=\tw@
    {\descriptionlabel{#1}}
    \hspace{\labelsep}%
  \else \ifnum\enit@type=\z@ \refstepcounter{\@listctr}\fi \quad\@itemlabel\hspace{\labelsep}%
  \fi}
\makeatother \parindent=0pt

\spnewtheorem{observation}{Observation}{\bfseries}{\itshape}
\usepackage[noline,noend,linesnumbered]{algorithm2e}
\usepackage{caption}

\newcommand{\oldcomment}[1]{}

\newcommand{\tool}[1]{\textsf{#1}}
\newcommand{\KoAT}[0]{\tool{KoAT}}
\newcommand{\braced}[1]{\left\lbrace #1 \right\rbrace}
\newcommand{\var}{\normalfont\texttt}

\newcommand{\eval}[2]{\llbracket #1 \rrbracket_{#2}}

\newcommand{\NN}{\mathbb{N}}
\newcommand{\ZZ}{\mathbb{Z}}

\newcommand{\RR}{\mathbb{R}}

\newcommand{\NNC}{\overline{\mathbb{N}}}

\newcommand{\landau}{\mathcal{O}}

\providecommand{\monus}{
  \mathbin{
    \vphantom{+}
    \text{
      \mathsurround=0pt 
      \ooalign{
        \noalign{\kern-.35ex}
        \hidewidth$\smash{\cdot}$\hidewidth\cr 
        \noalign{\kern.35ex}
        $-$\cr 
      }%
    }%
  }%
}

\newcommand{\FormulaSet}{\mathcal{F}}
\newcommand{\true}{\var{true}}

\newcommand{\initial}{\sigma_0}
\newcommand{\valuation}{\sigma}
\newcommand{\Valuation}{\Sigma}

\newcommand{\VSet}{\mathcal{V}}

\newcommand{\guard}{g}
\newcommand{\update}{\eta}
\newcommand{\Program}{\mathcal{P}}
\newcommand{\TSet}{\mathcal{T}}

\newcommand{\LSet}{\mathcal{L}}

\newcommand{\location}{\ell}
\newcommand{\locationInitial}{\location_0}

\newcommand{\IntProgram}{(\VSet,\LSet,\locationInitial,\TSet)}
\DeclareMathOperator{\rc}{rc}

\newcommand{\actt}{t}

\newcommand{\Size}{{\mathcal{SB}}}

\newcommand{\MRF}{\text{M}\Phi\text{RF}}
\newcommand{\MRFs}{\text{M}\Phi\text{RFs}}

\crefname{definition}{Def.}{Def.}
\crefname{example}{Ex.}{Ex.}
\crefname{counterexample}{Counterex.}{Counterex.}
\crefname{appendix}{App.}{App.}
\crefname{ex}{Ex.}{Ex.}
\crefname{theorem}{Thm.}{Thm.}
\crefname{thm}{Thm.}{Thm.}
\crefname{lemma}{Lemma}{Lemmas}
\crefname{lem}{Lemma}{Lemmas}
\crefname{remark}{Rem.}{Rem.}
\crefname{observation}{Observation}{Observations}
\crefname{section}{Sect.}{Sect.}
\crefname{subsection}{Sect.}{Sect.}
\crefname{subsubsection}{Sect.}{Sect.}
\crefname{line}{Line}{Lines}
\crefname{corollary}{Cor.}{Cor.}
\crefname{cor}{Cor.}{Cor.}
\crefname{figure}{Fig.}{Fig.}
\crefname{enumi}{}{}
\crefname{algorithm}{Alg.}{Alg.}
\crefname{observation}{Obs.}{Obs.}
\Crefname{observation}{Obs.}{Obs.}

\usepackage{marvosym}

\author{Nils Lommen$^{(\href{mailto:lommen@cs.rwth-aachen.de}{\mbox{\Letter}})}$\orcidID{0000-0003-3187-9217} \and Éléanore Meyer\orcidID{0000-0003-1038-4944} \and Jürgen Giesl\orcidID{0000-0003-0283-8520}}
\authorrunning{N.\ Lommen \and É.\ Meyer \and J.\ Giesl}
\institute{RWTH Aachen University, Aachen, Germany\\
  \email{\{lommen,eleanore.meyer,giesl\}@cs.rwth-aachen.de}}
\title{\tool{KoAT}: Automatic Complexity and Termination Analysis of Integer Programs\thanks{funded by the DFG Research Training Group 2236 UnRAVeL}}
\titlerunning{\tool{KoAT}: Automatic Complexity and Termination Analysis of Integer Programs}

\allowdisplaybreaks

\begin{document}
\maketitle \begin{abstract}
  \tool{KoAT} is a tool to automatically infer complexity bounds and prove termination of (possibly recursive) integer programs.
  To this end, \tool{KoAT} implements an alternating modular inference of upper runtime and size bounds for program parts.
  In particular, \tool{KoAT} uses a portfolio of different techniques to analyze subprograms.
  The power of our approach is demonstrated by an extensive experimental evaluation.
\end{abstract}

\section{Introduction}
Ensuring safety, security, and efficiency of software systems is a central goal of formal methods.
While functional verification is used to prove that programs compute correct results, it is also crucial to analyze their resource consumption, such as worst-case execution time and termination.
Since complex software is often abstracted into simplified integer programs during verification, inferring runtime and resource bounds for such integer programs is an important task in automated program analysis.

There exist numerous approaches to analyze complexity of programs on integers automatically, e.g., \cite{Ben-AmramGenaim17,albert2019ResourceAnalysisDriven,sinn2017ComplexityResourceBound,brockschmidt2016AnalyzingRuntimeSize,Flores-MontoyaH14,hoffmann2017AutomaticResourceBound,lopez18IntervalBasedResource,carbonneaux2015CompositionalCertifiedResource,albert2012CostAnalysisObjectoriented,pham2024RobustResourceBounds,hoffmann2012MultivariateAmortizedResource,albert2013InferenceResourceUsage,handley2019LiquidateYourAssets,lopez18IntervalBasedResource,albert2008AutomaticInferenceUpper,flores-montoya2016UpperLowerAmortized,frohn2022ProvingNonTerminationLower,DBLP:journals/mscs/HoffmannJ22,Dynaplex2021,DBLP:journals/tplp/RustenholzKCL24}.
We developed a \emph{modular} technique for complexity analysis of programs with built-in integers and implemented it in the complexity analysis tool \KoAT{}.
It automatically infers runtime bounds for integer programs possibly consisting of multiple loops by handling some subprograms as \emph{twn}-loops (``triangular weakly non-linear loops'', where there exist ``complete'' techniques for analyzing termination and complexity) \cite{lommen2022AutomaticComplexityAnalysis,lommen2023TargetingCompletenessUsing,lommen2024TargetingCompletenessComplexity,frohn2020TerminationPolynomialLoops,hark2020PolynomialLoopsTermination} and by using linear or multiphase-linear ranking functions ($\MRFs$) for other subprograms \cite{giesl2022ImprovingAutomaticComplexity,Ben-AmramGenaim17,Ben-AmramGenaim19,brockschmidt2016AnalyzingRuntimeSize,lommen2026FunctionCalls,ben-amram2014RankingFunctionsLinearConstraint,heizmann2015RankingTemplatesLinear,podelski2004CompleteMethodSynthesis}.
To lift local runtime bounds for isolated subprograms to bounds within the full program, we also infer size bounds on the values of the variables before entering such a subprogram.
By inferring bounds for one subprogram after the other, in the end we obtain a bound on the runtime of the whole program.
Moreover, besides its application for complexity analysis, we recently extended \KoAT{} by a specific mode in order to prove termination of integer programs.
While \KoAT{} can also analyze probabilistic programs \cite{DBLP:conf/tacas/MeyerHG21,lommenControlFlowRefinementComplexity2024}, here we focus on non-probabilistic programs to ease the presentation.
\begin{example}
  \label{exa:introduction}
  Consider the program in \Cref{fig:pseudocode} which consists of two nested \textbf{while}-loops.
  \begin{figure}[tb]
    \begin{align*}
       & \textbf{while } (x_1 > 0) \textbf{ do } \\
       & \qquad (x_1,x_2,x_3) \leftarrow (x_1 - 1, x_1, 1) \\
       & \qquad \textbf{while } (x_2 > x_3 \, \wedge \, x_2 > 0) \textbf{ do }
      (x_2,x_3) \leftarrow (2\cdot x_2, \,	3\cdot x_3)\textbf{ end } \\
       & \textbf{ end }
    \end{align*}
    \caption{Program with two Nested Loops}
    \label{fig:pseudocode}
  \end{figure}
  The outer \textbf{while}-loop is executed at most $x_1$ times.
  If one just considers the inner \textbf{while}-loop in isolation, then it is executed at most $\log_2(x_2) + 2$ times (see \Cref{sect:analysis}).
  This runtime bound can be obtained by our technique based on \emph{twn}-loops, but not by linear ranking functions.
  Note that $x_2$'s value is bounded by $x_1$ before executing the inner loop, where ``$x_1$'' refers to the initial value of the program variable.
  This observation is crucial to infer the (non-isolated) runtime of the inner loop since it is executed at most $x_1 \cdot (\log_2(\text{size}(x_2)) + 2) = x_1 \cdot (\log_2(x_1) + 2)$ times, where ``$\text{size}(x_2)$'' denotes an over-approximation of the (absolute) value of $x_2$ before entering the inner loop.
\end{example}

\paragraph{Structure.}
In \Cref{sect:ip}, we formalize our notion of integer programs.
Afterwards, we sketch the techniques implemented in \KoAT{} to analyze such integer programs in \Cref{sect:analysis}.
In \Cref{sect:relatedWork} we discuss related work.
We conclude with an experimental evaluation in \Cref{sect:evaluation}.

\section{Complexity and Termination Analysis of Integer Programs}
In this section, we present \KoAT's modular approach for the automatic analysis of runtime complexity and termination of integer programs.
More precisely, in \Cref{sect:ip}, we introduce our formalism of integer programs.
In \Cref{sect:analysis}, we then sketch the techniques used to analyze these programs.
\subsection{Integer Programs}
\label{sect:ip}
For integer programs, we use a formalism based on transitions.
As usual, $\ZZ[\VSet]$ denotes the polynomial ring with variables $\VSet$ and integer coefficients.
$\FormulaSet(\VSet)$ denotes the set of formulas, where a formula is a propositional combination of polynomial (in)equations over the variables $\VSet$ (i.e., expressions like $x + y \leq 2 \land x^2 = 2$).
\begin{definition}[Integer Program]
  \label{Integer Program}
  $\IntProgram$ is an \emph{integer program} where
  \begin{itemize}
    \item $\VSet$ is a finite set of (program) \emph{variables},
    \item $\LSet$ is a finite set of \emph{locations} with an \emph{initial location}
          $\location_0\in\LSet$, and
    \item $\TSet$ is a finite set of \emph{transitions}.
          A transition is a 4-tuple $(\location,\guard,\update,\location')$ with a \emph{start location} $\location\in\LSet$, \emph{target location} $\location'\in\LSet\setminus\braced{\location_0}$, \emph{guard} $\guard\in\FormulaSet(\VSet)$, and \emph{update function} $\update: \VSet\rightarrow\ZZ[\VSet]$ mapping variables to update polynomials.
  \end{itemize}
\end{definition}
An integer program may have two kinds of non-determinism.
Non-deterministic branching is realized by multiple transitions with the same start location whose guards are non-exclusive.
Non-deterministic sampling can be modeled by temporary variables (which can be restricted in the guard of a transition).
For simplicity, we omitted these temporary variables in \Cref{Integer Program} (see \cite{giesl2022ImprovingAutomaticComplexity,lommen2022AutomaticComplexityAnalysis,lommen2024TargetingCompletenessComplexity} for the extended definition).
Moreover, in \cite{lommen2026FunctionCalls} we extended our approach to integer programs with \emph{function calls}, including non-tail recursion.
\begin{example}
  In the integer program of \cref{fig:ITS} which corresponds to the pseudocode of \cref{fig:pseudocode}, we omitted identity updates $\update(v) = v$ and guards where $\guard$ is $\true$.
  Here, the variables are $\VSet = \braced{x_1,x_2,x_3}$, and $\LSet = \{\location_0, \location_1, \location_2\}$ are the locations, where $\location_0$ is the initial location.
  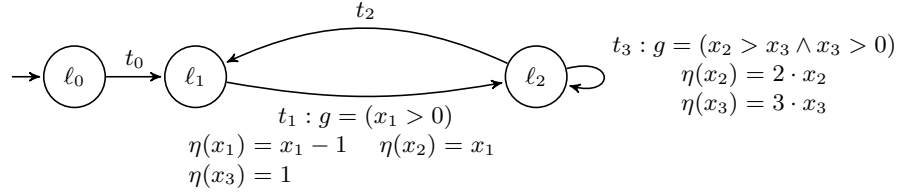
\begin{figure}[t]
    \centering
    \begin{tikzpicture}[->,>=stealth',shorten >=1pt,auto,node distance=3.5cm,semithick,initial text=$ $]
      \node[state,initial] (q0) {$\location_0$};
      \node[state] (q1) [right of=q0,xshift=-1.9cm]{$\location_1$};
      \node[state] (q2) [right of=q1, node distance=4.5cm]{$\location_2$};
      \draw (q0) edge node [text width=3.5cm,align=center,above] {\footnotesize $t_0$} (q1);
      \draw (q2) edge[bend left=-25] node [text width=4cm,align=center,above]
        {$t_2$}
      (q1);
      \draw (q2) edge [loop right] node [text width=3.75cm,align=center]
        {
          \footnotesize $t_3:\guard = (x_2 > x_3 \wedge x_3 > 0)$ \\
          $
            \begin{array}{rcl}
              \update(x_2) & = & 2\cdot x_2 \\
              \update(x_3) & = & 3\cdot x_3 \\
            \end{array}
          $
        }
      (q2);
      \draw (q1) edge[bend left=-10] node [text width=3cm, below, align=center]
        {
          \footnotesize \mbox{$t_1 : \guard = (x_1 > 0)$}
          \hspace*{-1cm}
          $
            \begin{array}{rclcrl}
              \update(x_1) & = & x_1 - 1 \quad & \update(x_2) & = & x_1 \\
              \update(x_3) & = & 1                                      \\
            \end{array}
          $
        }
      (q2);
    \end{tikzpicture}
    \caption{An Integer Program Corresponding to \Cref{fig:pseudocode}}\label{fig:ITS}
  \end{figure}
\end{example}

A \emph{state} $\valuation\colon \VSet\to \ZZ$ is a mapping from variables to numbers.
In the following, $\Valuation$ denotes the set of all states.
A \emph{configuration} is an element of $\LSet \times \Valuation$.
To define the semantics of integer programs, an evaluation step moves from one configuration $(\location,\valuation)\in\LSet\times\Valuation$ to another configuration $(\location',\valuation')$ via a transition $(\location, \guard, \update, \location')$ where the state $\valuation\in\Valuation$ satisfies the guard $\guard$ (denoted by $\valuation(\guard) = \normalfont{\true}$).
Let $\valuation(\update(v))$ denote the expression that results from replacing all variables $x$ by $\valuation(x)$ in the polynomial $\update(v)$.
Then $\valuation'$ is obtained by applying $\valuation$ on the update $\update$ (i.e., $\valuation'(v) = \valuation(\update(v))$ for all $v\in\VSet$).
From now on, we fix a program $\Program = \IntProgram$.

\begin{definition}[Evaluation of Integer Programs]
  \label{def:Evaluation of Integer Programs}
  A \emph{configuration} is an element of $\LSet \times \Valuation$.
  For two configurations $(\location,\valuation)$ and $(\location',\valuation')$, and a transition $t = (\location_t,\guard,\update,\location_{t}')\in\TSet$, $(\location,\valuation)\rightarrow_t(\location',\valuation')$ is an \emph{evaluation} step by $t$ if
  \begin{itemize}
    \item $\location = \location_t$ and $\location' = \location_{t}'$,
    \item $\valuation(\guard) = \normalfont{\true}$, and
    \item for every variable $v\in\VSet$ we have $\valuation'(v) = \valuation(\update(v))$.
  \end{itemize}
  We denote the union of all relations $\to_t$ for $t \in \TSet$ by $\to_{\TSet}$.
  Moreover, we abbreviate $(\location_0,\valuation_0)\rightarrow(\location_1,\valuation_1)\rightarrow \cdots \rightarrow(\location_k,\valuation_k)$ by $(\location_0,\valuation_0)\rightarrow^k(\location_k,\valuation_k)$ and write $(\location,\valuation)\rightarrow^*_\TSet(\location',\valuation')$ if $(\location,\valuation)\rightarrow^k_\TSet(\location',\valuation')$ for some $k\in\NN$.
\end{definition}

\begin{example}
  \label{exa:eval}
  When denoting states $\valuation\in\Valuation$ as tuples $(\valuation(x_1),\valuation(x_2),\valuation(x_3)) \in \ZZ^3$, then a run of \Cref{fig:ITS} that starts in $(5,0,0)$ has the form $(\location_0,(5,0,0)) \to_{t_0} (\location_1,(5,0,0)) \to_{t_1} (\location_2,(4,5,1)) \to_{t_3}^4 (\location_2,(4,80,81)) \to_{t_2} (\location_1,(4,80,81)) \to \cdots$.
\end{example}

For an initial state $\initial \in \Valuation$, the \emph{runtime complexity} $\rc(\initial)$ of a program corresponds to the length of a longest evaluation starting in $\initial$.

\begin{definition}[Runtime Complexity]
  \label{def:runtimeComplexityIntegerProg}
  The \emph{runtime complexity} is $\rc\colon\Valuation\rightarrow\NNC = \NN\cup\braced{\omega}$ with $\rc(\initial) = \sup\braced{k\in\NN\mid\exists (\location',\valuation').\, (\location_0,\initial)\rightarrow^k_\TSet(\location',\valuation')}$.
  An integer program is \emph{terminating} if and only if $\rc(\initial)\neq \omega$ for all $\initial \in \Valuation$.
\end{definition}

\begin{example}
  \label{exa:rc}
  For the program of \Cref{fig:ITS} and the state $\valuation\in\Valuation$ from \Cref{exa:eval} with $\valuation(x_1) = 5$ and $\valuation(x_2) = \valuation(x_3) = 0$, we obtain the runtime complexity $\rc(\valuation) = 25$.
\end{example}

Now we define our notion of \emph{bounds}.
\KoAT{} implements polynomial, logarithmic, and exponential bounds like $x^2 + y$, $\log_2(x)$, or $2^x$.
We only consider bounds which correspond to functions $f$ that are weakly monotonically increasing in all arguments, i.e., where $x \leq y$ implies $f(\ldots x \ldots) \leq f(\ldots y \ldots)$.
In this way, if $f$ and $g$ are both bounds, then $f\circ g$ is also a bound, i.e., bounds can be ``composed'' easily.
For example, we used this in the introductory \Cref{exa:introduction} where we inserted the size bound ``$\text{size}(x_2)$'' into the runtime bound $\log_2(x_2) + 2$ of the second loop.
In principle, every weakly monotonically increasing function could be used as a bound in our framework.
However, in \KoAT{} we restrict ourselves to bounds which are easy to represent and to compute with, and which cover the most prominent complexity classes.
\begin{definition}[Bounds]
  \label{def:bounds}
  The set of \emph{bounds} $\mathcal{B}$ is the smallest set containing~all
  \begin{itemize}
    \itemsep0.25em
    \item natural numbers $\NNC \subseteq \mathcal{B}$,
    \item variables $\VSet \subseteq \mathcal{B}$, and
    \item $\{b_1 + b_2, \max(b_1,b_2), b_1 \cdot b_2, p^{b_1}, \log_k(b_1)\} \subseteq \mathcal{B}$ for all $b_1,b_2 \in \mathcal{B}$, all polynomials $p\in\NN[\VSet]$, and all $k \in \RR_{> 1}$.\footnote{More precisely, instead of $\log_k(b_1)$ we use the function $\lceil\log_k(\max\braced{1,b_1})\rceil$ to ensure that bounds are well defined, weakly monotonically increasing, and evaluate to natural numbers.\label{rounding}
          }
  \end{itemize}
\end{definition}

We measure the size of variables by their absolute values.
For a state $\valuation\in\Valuation$, $|\valuation|$ is the state with $|\valuation|(v) =|\valuation(v)|$ for all $v \in \VSet$ and $\eval{b}{|\valuation|}$ denotes the evaluation of a bound $b \in \mathcal{B}$ at the state $|\valuation|$.
So if $\valuation(x) = 5$ and $\valuation(y) = -4$, then we have
\[
  \eval{x^2+y}{|\valuation|} = |\valuation|(x)^2 + |\valuation|(y) = 25 + 4 = 29.
\]

\medskip
A function $\mathcal{RB}: \TSet \rightarrow \mathcal{B}$ is a \emph{runtime bound} if for each transition $t$ and initial state $\initial \in \Valuation$, $\mathcal{RB}(t)$ evaluated in the state $|\initial|$ over-approximates the number of evaluation steps with $t$ in any run starting in the configuration $(\location_0, \valuation_0)$.
Let $\rightarrow^*_{\TSet} \circ \rightarrow_t$ denote the relation where arbitrary many evaluation steps are followed by a step with the transition $t$.
\begin{definition}[Runtime Bound]
  \label{def:rb}
  The function $\mathcal{RB}: \TSet \rightarrow \mathcal{B}$ is a \emph{runtime bound} if for all $t \in \TSet$ and all states $\initial \in \Valuation$ we have \[\eval{\mathcal{RB}(t)}{|\initial|} \; \geq \; \sup \braced{ k \in \NN \mid \exists \, (\location', \valuation').\; (\location_0, \valuation_0) \; (\rightarrow^*_{\TSet} \circ \rightarrow_t)^k \; (\location', \valuation') }.\]
\end{definition}

\begin{example}
  For the program of \Cref{fig:ITS}, a runtime bound is $\mathcal{RB}(t_0) = 1$ (as $t_0$ is not on a cycle), $\mathcal{RB}(t_1) = \mathcal{RB}(t_2) = x_1$ (this can be inferred by $\MRFs$), and $\mathcal{RB}(t_3) = x_1\cdot (\log_2(x_1) + 2)$ (this can be inferred by \emph{twn}-loops), see \Cref{sect:analysis}.
  Now a runtime bound for the full program is the sum of all these expressions.
  For example, we obtain $\eval{1 + x_1\cdot\log_2(x_1) + 4\cdot x_1}{|\valuation|} \approx 32.61 > 25 = \rc(\valuation)$ for the state $\valuation = (5,0,0)$ from \Cref{exa:eval,exa:rc}.
\end{example}

In the following, we define size bounds for variables $v$ after evaluating a transition $t\in\TSet$:
$\Size(t,v)$ is a \emph{size bound} for the variable $v$ w.r.t.\ the transition~$t$ if for any run starting in the initial state $\initial\in\Valuation$, $\eval{\Size(t, v)}{|\initial|}$ is greater or equal to the largest absolute value of $v$ after evaluating $t$.
\begin{definition}[Size Bounds]
  \label{sizebounds}
  A function $\Size: (\TSet \times \VSet) \rightarrow \mathcal{B}$ is a \emph{size bound} if for all $(t, v) \in \TSet \times \VSet$ and all states $\initial \in \Valuation$ we have \[\eval{\Size(t, v)}{|\initial|} \geq \sup \braced{ |\valuation'(v)| \mid \exists\, \location' \in \LSet.\; (\location_0, \initial) \; (\rightarrow^*_{\TSet} \circ \rightarrow_\actt) \; (\location', \valuation')}.\]
\end{definition}
\begin{example}
  \label{ex:sizeBoundLifting}
  For example, a size bound for $t_1$ and $x_2$ in \Cref{fig:ITS} is $\Size(t_1, x_2) = x_1$.
  Such size bounds are needed for our modular analysis, e.g., to handle nested loops.
  The transition $t_3$ (with the location $\location_2$) can be considered as a subprogram of the full integer program, i.e., it corresponds to the inner loop of the program in \Cref{fig:ITS}.
  An evaluation in the subprogram $\braced{t_3}$ can only make $t_3$-steps and a local runtime bound for this subprogram is $\log_2(x_2) + 2$ (see \Cref{sect:analysis}).

  To consider evaluations in the full program of \Cref{fig:ITS}, we lift \emph{local} to \emph{global} runtime bounds.
  To this end, we need size bounds.
  As in \Cref{def:rb}, a global runtime bound for $t_3$ over-approximates how often $t_3$ can be applied in a run of the full program.
  To obtain a global runtime bound from a local runtime bound in the subprogram $\braced{t_3}$, (i) we must consider how often the subprogram $\braced{t_3}$ is entered (at most $\mathcal{RB}(t_1)$ times) and (ii) we must consider the sizes of the variables upon entry into the subprogram (i.e., at most $\Size(t_1, v)$ for all $v \in \VSet$).
  So, Problem (i) is captured by the runtime bounds of the \emph{entry transitions} (in our case only $t_1$ is an entry transition of the subprogram $\braced{t_3}$).
  Problem (ii) is handled via size bounds for the entry transitions (i.e., by replacing $x_2$ with $\Size(t_1, x_2)$ in the local runtime bound $\log_2(x_2) + 2$).
  Thus, for our example, we obtain $$\mathcal{RB}(t_3) = \mathcal{RB}(t_1)\cdot(\log_2(\Size(t_1, x_2)) + 2) = x_1\cdot (\log_2(x_1) + 2).$$
\end{example}

\subsection{Analyzing Integer Programs}
\label{sect:analysis}
\begin{figure}[tb]
  \begin{center}
    \begin{tikzpicture}[bullet/.style={circle, fill, inner sep=1pt}]

      \node[align=left] (koat) {\textbf{\textcolor{burgundy}{\Large \KoAT{}}} (Input: Integer Program as in \Cref{Integer Program})};

      \node[below=0.15cm of koat.south west, align=left, anchor=north west, shape=rectangle, rounded corners, draw=black, minimum height=4mm, minimum width=20mm, fill=gray!20] (left_text) {
        \textbf{Modular Runtime Bounds} \\[1mm]
        $\bullet$ $\MRFs$ \cite{giesl2022ImprovingAutomaticComplexity,Ben-AmramGenaim17,Ben-AmramGenaim19,brockschmidt2016AnalyzingRuntimeSize} \\
        $\bullet$ \emph{twn}-Loops
        \cite{lommen2022AutomaticComplexityAnalysis,hark2020PolynomialLoopsTermination,frohn2020TerminationPolynomialLoops,lommen2024TargetingCompletenessComplexity}
      };

      \node[right=1.2cm of left_text.north east,anchor=north west,align=left, shape=rectangle, rounded corners, draw=black, minimum height=4mm, minimum width=20mm, fill=gray!20
      ] (right_text) {
        \textbf{Modular Size Bounds} \\[1mm]
        $\bullet$ Accumulating Local Change \cite{brockschmidt2016AnalyzingRuntimeSize,lommen2026FunctionCalls} \\
        $\bullet$ Solvable Loops \cite{lommen2023TargetingCompletenessUsing,lommen2024TargetingCompletenessComplexity}
      };
      \node[font=\LARGE] (plus) at ($(left_text.east)!0.5!(right_text.west)$) {\textbf{$\rightleftharpoons$}};

      \node[below= 0.7cm of plus,shape=rectangle, rounded corners, draw=black, minimum height=4mm, minimum width=20mm, fill=gray!20] (cfr) {\textbf{CFR} \cite{domenech2018IRankFinder,domenech2019ControlFlowRefinementPartial,giesl2022ImprovingAutomaticComplexity,lommenControlFlowRefinementComplexity2024}};

      \begin{scope}[on background layer]
        \node[draw, thick, rounded corners=15pt, inner sep=8pt, fit=(koat) (left_text) (right_text) (cfr), fill=gray!10] (box) {};
      \end{scope}

      \node[below=1.7cm of left_text,shape=rectangle, rounded corners, draw=black, minimum height=4mm, minimum width=20mm, fill=blue!10] (termination) {\textbf{\textcolor{OliveGreen}{Termination}}};
      \node[right=1cm of termination,shape=rectangle, rounded corners, draw=black, minimum height=4mm, minimum width=20mm, fill=blue!10] (runtime_bounds) {\textbf{\textcolor{OliveGreen}{Runtime Bound}} (e.g., $x^2 + y$, $\log_2(x)$, or $2^x$)};

      \draw[<->, thick, dashed, shorten <=2pt, shorten >=2pt] (cfr.west) to[bend left=20, looseness=1.2] (left_text.south);

      \draw[<->, thick, dashed, shorten <=2pt, shorten >=2pt] (cfr.east) to[bend right=20, looseness=1.2] (right_text.south);
      \draw[->, thick] (box.south -| left_text) -- (termination);
      \draw[->, thick] (runtime_bounds.north |- box.south) -- (runtime_bounds.north);
    \end{tikzpicture}
  \end{center}
  \caption{\KoAT{}: Automatic Complexity and Termination Analysis of Integer Programs}\label{fig:koat}
\end{figure}
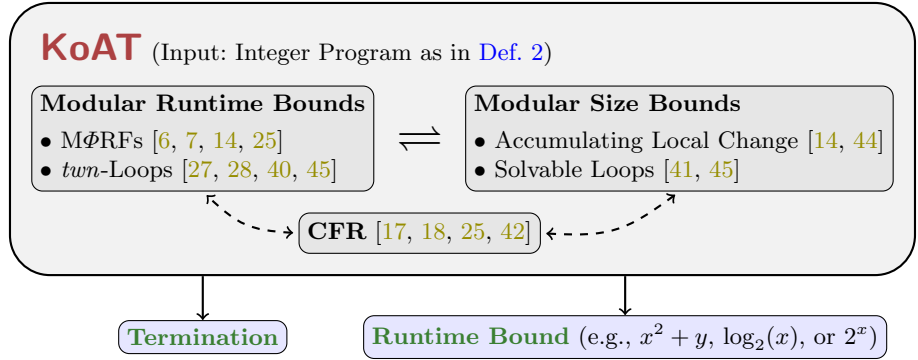
We now sketch our modular approach to analyze integer programs.
An overview of our tool \KoAT{} is depicted in \Cref{fig:koat}.
In order to compute runtime and size bounds, we analyze subprograms in topological order, i.e., in case of multiple consecutive loops, we start with the first loop and propagate knowledge about the resulting values of variables to subsequent loops.
By handling one subprogram after the other, in the end we obtain a bound on the runtime complexity of the whole program.
The approach uses size bounds to infer runtime bounds, i.e., size bounds on the entry transitions of subprograms are used to lift local runtime bounds for isolated subprograms to global runtime bounds for the full program. Similarly, we also use runtime bounds to infer size bounds, since runtime bounds provide information on how often the local change of a transition can be performed.
Hence, runtime and size bound computations are alternated until all bounds are finite or no bound can be improved further.
Finally, if our analysis fails to infer a ``good enough'' runtime bound, then \KoAT{} applies \emph{control-flow refinement} (CFR) to the integer program \cite{lommenControlFlowRefinementComplexity2024,giesl2022ImprovingAutomaticComplexity,domenech2018IRankFinder,domenech2019ControlFlowRefinementPartial} in order to obtain a program that is easier to analyze.
In the following, we describe \KoAT's techniques to infer local runtime bounds (via $\MRFs$ or \emph{twn}-loops), size bounds, and CFR in more detail.

\paragraph{Multiphase-Linear Ranking Functions ($\MRFs$).}
\KoAT{} uses individual ranking functions to infer local runtime bounds for different subprograms.
A ranking function assigns an integer to each configuration $(\ell,\valuation) \in \LSet \times \Valuation$.
The value of a ranking function must not increase in evaluation steps, and it has to decrease by at least one in every evaluation step where a specific transition is applied.
Moreover, it must have a positive value before each such evaluation step.
\begin{example}
  For the integer program in \Cref{fig:ITS}, \KoAT{} uses $\MRFs$ to infer runtime bounds for the transitions of the outer loop.
  For example, we can infer a runtime bound for $t_1$ by considering the ranking function
  \[
    f(\ell_0,\valuation) = f(\ell_1,\valuation) = f(\ell_2, \valuation) = \valuation(x_1).
  \]
  The reason is that $t_1$ decreases this ranking function while the other transitions do not increase it.
  In contrast, $\MRFs$ cannot be used to obtain a runtime bound for the inner loop consisting just of transition $t_3$. Instead, this inner loop is analyzed by our technique for \emph{twn}-loops.
  As discussed in \Cref{ex:sizeBoundLifting}, the local runtime bound for this inner \emph{twn}-loop is then lifted to a global runtime bound for $t_3$ in the full program.
\end{example}

A (nested) $\MRF$ \cite{Ben-AmramGenaim17,Ben-AmramGenaim19} extends the idea of ranking functions and uses a mapping $f_i$ for every ``phase'' $1 \leq i \leq d$ of a program.
Before each evaluation step, the value of the last phase $f_d$ is required to be positive.
Moreover, for all~$i$, the sum of $f_{i - 1}$ and $f_i$ before the evaluation step must be larger than the function $f_i$ after the update.
We set $f_0 (\ell,\valuation)$ to $0$ for all configurations $(\ell,\valuation)$.
Thus, $f_1$ must be decreasing with each update.
If $f_1$ becomes negative, then
\[
  f_1(\ell,\valuation) + f_2(\ell,\valuation) < f_2(\ell,\valuation)
\]
and thus, now $f_2$ has to be decreasing with every update, and so on until $f_d$ becomes decreasing.
The program eventually terminates, since $f_d$ must take a positive number whenever the program can be executed further.
Moreover, if all functions $f_1,\ldots,f_d$ are defined via linear polynomials, then this implies a linear runtime bound for the subprogram. We refer to \cite{giesl2022ImprovingAutomaticComplexity} for further details and the formal definition of $\MRFs$ for subprograms.

\KoAT{} only implements ranking functions via linear polynomials since these are easy to generate automatically \cite{podelski2004CompleteMethodSynthesis}, and since they are already quite powerful when combining them within $\MRFs$.
Moreover, in contrast to \emph{twn}-loops, $\MRFs$ are applicable to arbitrary integer programs with linear arithmetic.

\paragraph{Triangular Weakly Non-Linear-Loops.}
As an alternative approach to infer local runtime bounds, we integrated a technique \cite{frohn2020TerminationPolynomialLoops,hark2020PolynomialLoopsTermination} to analyze termination of \emph{triangular weakly non-linear loops} (\emph{twn}-loops) in \KoAT{} \cite{lommen2022AutomaticComplexityAnalysis,lommen2024TargetingCompletenessComplexity,lommen2023TargetingCompletenessUsing}.
This approach also allows us to analyze programs with \emph{non-linear} arithmetic, while the automated generation of $\MRFs$ is essentially limited to programs with linear arithmetic.
An example for a terminating \emph{twn}-loop which corresponds to the subprogram~$\braced{t_3}$ from \Cref{fig:ITS} is given by:
\begin{equation}
  \label{WhileExample}
  \textbf{while } (x_2 > x_3 \, \wedge \, x_3 > 0) \textbf{ do }
  (x_2,x_3) \leftarrow (2\cdot x_2, \,	3\cdot x_3) \quad \textbf{ end }
\end{equation}
This loop does not admit a M$\Phi$RF over $\mathbb{R}$ (see \cite{heizmann2015RankingTemplatesLinear}).
The guard of such \emph{twn}-loops are propositional formulas over (possibly non-linear) polynomial inequations.
The update is \emph{triangular}, i.e., we can order the variables such that the update of any $x_i$ does not depend on the variables $x_1, \ldots, x_{i-1}$ with smaller indices.
So the restriction to triangular updates prohibits ``cyclic dependencies'' of variables (e.g., where the new values of $x_1$ and $x_2$ both depend on the old values of $x_1$ and $x_2$).
From a practical point of view, the restriction to triangular loops seems quite natural.
For example, in \cite{frohn2022CalculusModularLoop}, $1511$ polynomial loops were extracted from the \emph{Termination Problems Data Base} (TPDB) \cite{tpdb}, the benchmark collection which is used at the annual \emph{Termination and Complexity Competition} (TermComp) \cite{termcomp}, and only $26$ of them were non-triangular.
Furthermore, the update of a \emph{twn}-loop is \emph{weakly non-linear}, i.e., no variable $x_i$ has a non-linear occurrence in its own update.
With triangularity and weak non-linearity, one can compute a \emph{closed form} which corresponds to applying the loop's update~$n$ times and which is suitable for (dis)proving termination and inferring runtime bounds.
Using these closed forms, termination can be reduced to a formula over~$\mathbb{Z}$~\cite{frohn2020TerminationPolynomialLoops} (whose satisfiability is decidable for linear arithmetic and where SMT solvers often also prove (un)satisfiability in the non-linear case).
Furthermore, the closed forms can be used to infer runtime bounds.
More precisely, one can always compute a polynomial runtime bound for every terminating \emph{twn}-loop.
Moreover, these bounds are linear if the loop only contains linear arithmetic.
So for linear \emph{twn}-loops, termination is decidable and a linear runtime bound can always be computed (if the loop is terminating).

The bounds are even logarithmic if in addition, the coefficients of the variables in their own updates have pairwise different absolute values \cite{lommen2024TargetingCompletenessComplexity}.
This is the case for the \emph{twn}-loop \eqref{WhileExample}, because the coefficients of $x_2$ and $x_3$ in their own updates~$\update(x_2)$ and $\update(x_3)$ are $2$ and $3$, which are different (absolute) values.
Thus, here we can infer the logarithmic runtime bound $\log_2(x_2) + 2$ \cite[App.\ C]{lommen2026FunctionCalls}.
An analogous criterion for logarithmic runtimes can also be obtained for non-linear \emph{twn}-loops.
For future work, it would be interesting to develop classes of loops where one can give tightness guarantees for the obtained runtime bounds.

\paragraph{Size Bounds.}
As outlined before, size bounds are an important ingredient of our modular approach for complexity analysis.
Currently, we use two techniques to infer size bounds in \KoAT{}.
On the one hand, we integrated a procedure for \emph{solvable} loops \cite{DBLP:conf/tacas/Kovacs08,kovacs2006CombiningLogicAlgebraic}.
For all loops from this subclass (which includes all \emph{twn}-loops), whenever we have a runtime bound, the procedure computes finite size bounds.
To this end, we over-approximate the effect of a loop by computing closed forms and instantiating the loop counter by the runtime bound.
On the other hand, we compute size bounds by considering the local change resulting from a transition.
To compute the accumulated change that results from the repeated execution of the
transition, we use its runtime bound
\cite{brockschmidt2016AnalyzingRuntimeSize,lommen2024TargetingCompletenessComplexity,lommen2023TargetingCompletenessUsing}.
This approach is applicable to arbitrary programs and not restricted to loops.
Note that \KoAT{} only needs size bounds for complexity, but not for termination analysis.

\paragraph{Control-Flow Refinement.}
\begin{figure}[t]
  \centering
  \begin{minipage}[t]{0.3\linewidth}
    \LinesNotNumbered
    \begin{algorithm}[H]
      \While{$0 < x$}{
        \eIf {$y < z$}{
          $y \gets y + x$
        }{
          $x \gets x - 1$
        }
      }
    \end{algorithm}
    \captionof{figure}{Original Loop}\label{fig:crf}
  \end{minipage}
  \hspace*{.8cm}
  \begin{minipage}[t]{0.5\linewidth}
    \LinesNotNumbered
    \begin{algorithm}[H]
      \hspace*{.5cm}	\While{$0 < x \wedge y < z$}{
        \hspace*{.5cm}	$y \gets y + x$
      }
      \hspace*{.5cm}	\While{$0 < x \wedge y \geq z$}{
        \hspace*{.5cm}	$x \gets x - 1$
      }
    \end{algorithm}
    \hspace*{1cm}\\
    \captionof{figure}{After Control-Flow Refinement}\label{fig:unrolled}
  \end{minipage}
\end{figure}
In addition, we integrated \emph{control-flow refinement} \cite{domenech2018IRankFinder,domenech2019ControlFlowRefinementPartial} into our tool \KoAT{} \cite{giesl2022ImprovingAutomaticComplexity,lommenControlFlowRefinementComplexity2024}.
The main idea of CFR is to gain information on the values of variables and to sort out certain paths in the program.
For example, our adaption of the CFR technique from \cite{domenech2019ControlFlowRefinementPartial} in \KoAT{} detects that in \Cref{fig:crf}, after reaching a loop iteration where the \textbf{else}-case applies, one never reaches another loop iteration where the \textbf{if}-case is used.
Therefore, it transforms the integer program corresponding to \cref{fig:crf} into the program corresponding to \cref{fig:unrolled}.
While the programs are equivalent, the program in \cref{fig:unrolled} is easier to analyze, since here the two consecutive loops do not interfere with each other: $x$ and $z$ are constants in its first loop, while $y$ and $z$ are constants in its second~loop.

\section{Related Work}
\label{sect:relatedWork}
As mentioned in the introduction, there exist many approaches to analyze complexity of integer programs automatically, e.g., \cite{Ben-AmramGenaim17,albert2019ResourceAnalysisDriven,sinn2017ComplexityResourceBound,brockschmidt2016AnalyzingRuntimeSize,Flores-MontoyaH14,hoffmann2017AutomaticResourceBound,lopez18IntervalBasedResource,carbonneaux2015CompositionalCertifiedResource,albert2012CostAnalysisObjectoriented,pham2024RobustResourceBounds,hoffmann2012MultivariateAmortizedResource,albert2013InferenceResourceUsage,handley2019LiquidateYourAssets,lopez18IntervalBasedResource,albert2008AutomaticInferenceUpper,flores-montoya2016UpperLowerAmortized,frohn2022ProvingNonTerminationLower,DBLP:journals/mscs/HoffmannJ22,Dynaplex2021,DBLP:journals/tplp/RustenholzKCL24}.
Instead of representing integer programs via transitions, there are also techniques based on \emph{cost equation systems}, implemented in the tool \tool{CoFloCo} \cite{Flores-MontoyaH14,flores-montoya2016UpperLowerAmortized}.
This approach analyzes program parts independently and uses linear invariants to compose the results, i.e., it differs significantly from our approach which can also infer non-linear size bounds.
Similarly, in the tool \tool{PUBS} \cite{albert2008AutomaticInferenceUpper,albert2013InferenceResourceUsage}, \emph{cost relations} are analyzed which are a system of recursive equations that capture the cost of the program.
There are also numerous approaches for automatic resource analysis of functional programs, often based on amortized analysis (see \cite{DBLP:journals/mscs/HoffmannJ22} for an overview).
For example, an approach for automatic complexity analysis of \tool{OCaml} programs is presented in \cite{hoffmann2012MultivariateAmortizedResource,hoffmann2017AutomaticResourceBound}, which however has limitations w.r.t.\ modularity, see \cite{FroCoS17}, and is restricted to polynomial bounds.
There are also several approaches based on types, e.g., the resource consumption of \tool{Liquid Haskell} programs is encoded in a type system in \cite{handley2019LiquidateYourAssets}, but here bounds are not inferred automatically.
Another line of work automatically infers bounds by generating and solving recurrence relations, e.g., \cite{Dynaplex2021,DBLP:journals/tplp/RustenholzKCL24}.

There also exist tools which analyze the runtime complexity of \tool{C}-code, e.g., \tool{Loopus} \cite{sinn2017ComplexityResourceBound} or \tool{MaxCore} \cite{albert2019ResourceAnalysisDriven} with \tool{CoFloCo} or \tool{PUBS} in the backend.
For \tool{KoAT}, we use \tool{Clang} \cite{clang} and \tool{llvm2kittel} \cite{falke2011TerminationAnalysisPrograms} to transform pointer-free \tool{C} programs into integer transition systems.
To prove termination of more general \tool{C} programs, we developed the framework \tool{AProVE (KoAT + LoAT)} \cite{lommen2025AProVEKoATLoAT}, which participates in the annual \emph{Software Verification Competition}
(SV-COMP) \cite{svcomp}.
In addition, \tool{AProVE} also uses \tool{KoAT} to infer runtime bounds for more general \tool{C} programs~\cite{DBLP:journals/jlp/HenselGFS18}.
Moreover, \tool{KoAT} is used as a backend in the tool \tool{Pico} for cost analysis of GPU kernels \cite{blike2026ModularStaticCost}.

There is a wealth of work on automated termination analysis of integer programs.
Most approaches are based on variants of ranking functions (e.g., \cite{podelski2004CompleteMethodSynthesis,heizmann2015RankingTemplatesLinear,ben-amram2014RankingFunctionsLinearConstraint,Ben-AmramGenaim17,Ben-AmramGenaim19}).
Moreover, there are also classes of programs where termination is decidable (e.g., \cite{frohn2020TerminationPolynomialLoops,braverman2006TerminationIntegerLinear,tiwari2004TerminationLinearPrograms,xu2013SymbolicTerminationAnalysis,hosseini2019TerminationLinearLoops}).
There exist numerous tools which can automatically prove or disprove termination of integer programs, e.g., \tool{Golem} \cite{blicha2025GolemFlexibleEfficient}, \tool{iRankFinder} \cite{domenech2018IRankFinder}, \tool{LoAT} \cite{frohn2022ProvingNonTerminationLower},\linebreak[2]
\tool{MuVal}~\cite{unnoModularPrimalDualFixpoint2023}, \tool{T2}
\cite{brockschmidt2016T2TemporalProperty}, and \tool{VeryMax}
\cite{borralleras2017ProvingTerminationConditional} which participated in last year's TermComp.\footnote{In recent years, also several data-driven techniques for termination analysis were developed which performed successfully at competitions like SV-COMP \cite{svcomp}, e.g., \cite{Verifuzz,Proton}.
  However, in general these techniques can yield unsound results.}
The tools \tool{Golem} and \tool{LoAT} are specialized in \emph{disproving} termination, whereas \KoAT{} can only \emph{prove} termination.
Moreover, \tool{LoAT} infers lower runtime bounds.
Thus, by running \tool{LoAT} in parallel with \KoAT{} one can infer tight asymptotic bounds whenever the bounds of both tools match.

\section{Evaluation and Conclusion}
\label{sect:evaluation}
We presented the tool \KoAT{} for termination and complexity analysis of integer programs.
While the first version of \KoAT{} dates back to \cite{brockschmidt2016AnalyzingRuntimeSize}, this version did not have a specialized mode for termination analysis, it only applied classical ranking functions for the computation of runtime bounds (but no $\MRFs$), and it did not use any techniques based on \emph{twn}- or solvable loops.
Thus, we re-implemented \KoAT{} completely and integrated the results of our papers \cite{lommen2022AutomaticComplexityAnalysis,giesl2022ImprovingAutomaticComplexity,lommen2023TargetingCompletenessUsing,lommen2026FunctionCalls,lommen2024TargetingCompletenessComplexity}.

Moreover, our new implementation of \KoAT{} can also analyze \emph{expected} runtimes of \emph{probabilistic} integer programs \cite{DBLP:conf/tacas/MeyerHG21,lommenControlFlowRefinementComplexity2024}.
To this end, we lifted our modular framework to the probabilistic setting, leading to the notions of \emph{expected} runtime and size bounds.
They over-approximate the expected number of evaluation steps with a specific transition and the expected absolute value of a variable after evaluating a specific transition.
In general, these expected values may depend on each other, such that it is not always possible to combine already computed expected runtime and size bounds to derive new, improved bounds.
In such cases, \KoAT{} falls back to the computation of \emph{non-probabilistic} bounds by over-approximating all probabilistic with non-probabilistic behavior.
The obtained non-probabilistic bounds are then integrated into the analysis of the original probabilistic program.
Thus, any improvement in the analysis of non-probabilistic programs can also be used when analyzing probabilistic ones.

\KoAT{} is an open-source tool written in \tool{OCaml}.
In the beginning of the analysis, \KoAT{} preprocesses the program, e.g., by extending the guards of transitions with polyhedron invariants inferred by \tool{Apron} \cite{jeannet2009ApronLibraryNumerical}.
For all SMT problems (e.g., to infer $\MRFs$ and to prove termination of \emph{twn}-loops), \KoAT{} calls \tool{Z3} \cite{moura2008Z3SMTSolver}.

To evaluate the power of \KoAT{}, we used the benchmarks from the TPDB in the categories \tool{Complexity\_ITS} and \tool{Termination\_ITS} for integer transition systems, which are used in the annual TermComp.
For the (similar) results of a corresponding evaluation on \textsf{C} programs see \cite{lommen2024TargetingCompletenessComplexity}.
In our evaluation, we compare \KoAT{} with all tools which participated in the respective categories at TermComp.
Moreover, we compared \KoAT{} to the original \KoAT{} implementation of~\cite{brockschmidt2016AnalyzingRuntimeSize}.
To distinguish this original implementation from our re-imple\-mentation, we refer to the tool of \cite{brockschmidt2016AnalyzingRuntimeSize} as \tool{KoAT1} in the following.
As mentioned, \tool{KoAT1} does not have a specific procedure for termination analysis.
For complexity analysis, we evaluated our novel version of \KoAT{} against the tools \tool{KoAT1} and \tool{CoFloCo}.
We ran \KoAT{} using its most powerful configuration, in which all our recent improvements are enabled (i.e., $\MRFs$, runtime bounds for \emph{twn}-loops, size bounds for solvable loops, and control-flow refinement).
\begin{table}[t]
  \caption{Evaluation for Complexity Analysis (\tool{Complexity\_ITS}, 838 Benchmarks)}
  \label{fig:Complexity}
  \renewcommand{\arraystretch}{1.1}
  \setlength{\tabcolsep}{2.75pt}
  \begin{center}
    \begin{tabular}{lcccccccc}
      \toprule       & $\landau(1)$ & $\landau(\log(n))$ & $\landau(n)$ & $\landau(n^2)$ & $\landau(n^{>2})$ & $<\omega$ & $\mathrm{AVG^+(s)}$ & $\mathrm{AVG(s)}$ \\
      \cmidrule(lr){2-2}
      \cmidrule(lr){3-3}
      \cmidrule(lr){4-4}
      \cmidrule(lr){5-5}
      \cmidrule(lr){6-6}
      \cmidrule(lr){7-7}
      \cmidrule(lr){8-8}
      \cmidrule(lr){9-9}
      \KoAT{}        & 132          & 9                  & 261          & 113            & 24                & 548       & 2.81                & 7.17              \\
      \tool{KoAT1}   & 132          & 0                  & 216          & 104            & 14                & 475       & 0.67                & 3.33              \\
      \tool{CoFloCo} & 125          & 0                  & 230          & 93             & 9                 & 457       & 1.50                & 5.38              \\
      \bottomrule
    \end{tabular}
  \end{center}
\end{table}
\Cref{fig:Complexity} shows the results of our evaluation on \tool{Complexity\_ITS}, where as in last year's TermComp, we used a timeout of one minute per example.
All tools were run inside an Ubuntu Docker container on a machine with an AMD Ryzen 7 3700X octa-core CPU and $8 \, \mathrm{GB}$ of RAM.
The runtime bounds inferred by the tools are compared asymptotically as functions which depend on the largest initial absolute value $n$ of all program variables.
So for example, \KoAT{} proved an (at most) linear runtime bound for $132 + 9 + 261 = 402$ benchmarks.
So for these examples \KoAT{} can show that $\rc(\initial)\in\landau(n)$ for all initial states $\initial \in \Valuation$ where $|\initial|(v) \leq n$ for all $v \in \VSet$.
Overall, this configuration succeeds on $548$ examples, i.e., ``$< \omega$'' is the number of examples where a finite bound on the runtime complexity could be computed by the tool within the time limit.
``$\mathrm{AVG^+(s)}$'' denotes the average runtime of successful runs in seconds, whereas ``$\mathrm{AVG(s)}$'' is the average runtime of all runs.
While \KoAT{} is a bit slower, the table clearly shows that it is considerably more powerful than the two other tools for complexity analysis.
\begin{table}[t]
  \caption{Evaluation for Termination Analysis (\tool{Termination\_ITS}, 1222 Benchmarks)}
  \label{fig:Termination}
  \renewcommand{\arraystretch}{1.1}
  \setlength{\tabcolsep}{2.75pt}
  \begin{center}
    \begin{tabular}{lccccccc}
      \toprule \textbf{Result}              & \tool{T2}     & \tool{iRankFinder} & \tool{VeryMax} & \tool{MuVal} & \tool{KoAT}  & \tool{LoAT}  & \tool{Golem} \\
      \cmidrule(lr){1-1}
      \cmidrule(lr){2-2}
      \cmidrule(lr){3-3}
      \cmidrule(lr){4-4}
      \cmidrule(lr){5-5}
      \cmidrule(lr){6-6}
      \cmidrule(lr){7-7}
      \cmidrule(lr){8-8}
      \textbf{Terminating}                  & 606           & 632                & 627            & 515          & \textbf{635} & 2            & 0            \\
      \textbf{Non-Terminating}              & 422           & 385                & 364            & 448          & 0            & \textbf{522} & 350          \\
      \midrule \textbf{Total}               & \textbf{1028} & 1017               & 991            & 963          & 635          & 524          & 350          \\
      \midrule \textbf{$\mathrm{AVG^+(s)}$} & 2.03          & 5.45               & 6.67           & 3.96         & 1.96         & 1.38         & 2.42         \\
      \textbf{$\mathrm{AVG(s)}$}            & 7.08          & 12.99              & 14.51          & 15.51        & 7.94         & 11.24        & 9.87         \\
      \bottomrule
    \end{tabular}
  \end{center}
\end{table}

For termination analysis, we compared \KoAT{} against the tools which participated in last year's TermComp: \tool{Golem}, \tool{iRankFinder}, \tool{LoAT}, \tool{MuVal}, \tool{T2}, and \tool{VeryMax} on \tool{Termination\_ITS}, see \Cref{fig:Termination}.
In our experiments, \KoAT{} proved termination for 635 benchmarks in an average runtime of 1.96 seconds.
This shows that \KoAT{} is a powerful termination tool even though it is primarily designed for complexity analysis.
More precisely, in our experiments, \KoAT{} was the most powerful tool for \emph{proving} termination of programs, whereas \tool{LoAT} was the most powerful tool for \emph{disproving} termination, which proved non-termination on 522 benchmarks.
Overall, \tool{T2} solved the most examples with 1028 instances.

\KoAT's source code, a binary, and a Docker image are available at our webpage:
\[\mbox{\url{https://koat.verify.rwth-aachen.de/}}\]
This website also contains details on our input format and a \emph{web interface} to run different configurations of \KoAT{} directly online.
The detailed results of our evaluation can be found at \url{https://koat.verify.rwth-aachen.de/evaluation} \cite{webpage}.
\begin{credits}
  \subsubsection{\discintname}
  The authors have no competing interests to declare that are relevant to the content of this article.
  \subsubsection{Data Availability Statement.}
  An artifact with \KoAT's binary and Docker images in order to reproduce our experiments is available at \cite{artifact}:

  \[
    \mbox{\url{https://doi.org/10.5281/zenodo.18399478}}
  \]
\end{credits}

\printbibliography

\end{document}